\documentclass[australian,english,prl, singlespace, twocolumn]{revtex4-1}
\usepackage[T1]{fontenc}
\usepackage[latin9]{inputenc}
\usepackage{color}
\usepackage{amsmath}
\usepackage{amssymb}
\usepackage{graphicx}

\makeatletter
\@ifundefined{definecolor}
 {\usepackage{color}}{}
\makeatother

\makeatother

\usepackage{babel}
\begin{document}
\title{Testing macroscopic local realism using time-settings}
\author{M. Thenabadu$^{1}$, G-L. Cheng$^{1}$, T. L. H. Pham$^{2}$, L. V.
Drummond$^{2}$, L. Rosales-Z\'arate$^{3}$ and M. D. Reid$^{1,4}$}
\affiliation{$^{1}$Centre for Quantum and Optical Science Swinburne University
of Technology, Melbourne, Australia}
\affiliation{\selectlanguage{australian}%
$^{2}$\foreignlanguage{english}{Physics Department, University of
Melbourne, Melbourne, Australia}}
\affiliation{$^{3}$Centro de Investigaciones en \'Optica A.C., Le\'on, Guanajuato
37150, M\'exico}
\affiliation{\selectlanguage{english}%
$^{4}$Institute of Theoretical Atomic, Molecular and Optical Physics
(ITAMP),Harvard University, Cambridge, Massachusetts, USA}
\begin{abstract}
We show how one may test macroscopic local realism where, different
from conventional Bell tests, all relevant measurements need only
distinguish between two macroscopically distinct states of the system
being measured. Here, measurements give macroscopically distinguishable
outcomes for a system observable and do not resolve microscopic properties
(of order $\hbar$). Macroscopic local realism assumes: (1) macroscopic
realism (the system prior to measurement is in a state which will
lead to just one of the macroscopically distinguishable outcomes)
and (2) macroscopic locality (a measurement on a system at one location
cannot affect the macroscopic outcome of the measurement on a system
at another location, if the measurement events are spacelike separated).
To obtain a quantifiable test, we define $M$-scopic local realism
where the outcomes are separated by an amount $\sim M$. We first
show for $N$ up to $20$ that $N$-scopic Bell violations are predicted
for entangled superpositions of $N$ bosons (at each of two sites).
Secondly, we show violation of $M$-scopic local realism for entangled
superpositions of coherent states of amplitude $\alpha$, for arbitrarily
large $M=\alpha$. In both cases, the systems evolve dynamically
according to a local nonlinear interaction. The first uses nonlinear
beam splitters realised through nonlinear Josephson interactions;
the second is based on nonlinear Kerr interactions. To achieve
the Bell violations, the traditional choice between two spin measurement
settings is replaced by a choice between different times of evolution
at each site.
\end{abstract}
\maketitle
Motivated by Schrodinger's cat paradox \cite{s-cat}, much effort
has been devoted to testing quantum mechanics at a macroscopic level.
Quantum superpositions of macroscopically distinguishable states,
so-called cat-states, have been created in a number of different physical
scenarios \cite{supmat}. However, Leggett and Garg pointed out that
a very strong test of macroscopic quantum mechanics would give a method
to falsify \emph{all} possible alternative theories satisfying the
notion of macroscopic realism \cite{legggarg}.

To address this problem, Leggett and Garg formulated inequalities
\cite{Bell-2}, which if violated falsify a form of macroscopic realism
now called macro-realism \cite{legggarg,emary-review}. Leggett and
Garg's macro-realism combines two classical premises: The first premise
is \emph{macroscopic realism }(MR): For a system which has two macroscopically
distinguishable states available to it, as identifiable by a measurement
which gives one of two macroscopically distinguishable outcomes, the
system must at any time be in one or other of these states i.e. it
must be in a state which will lead to just one of the distinct outcomes.
Macroscopic realism implies the existence of a hidden variable to
predetermine outcomes of measurements that are macroscopically distinct
\cite{legggarg}. In Schrodinger's paradox, the assumption of macroscopic
realism is that Schrodinger's cat is always dead \emph{or} alive,
prior to any measurement.

The second Leggett-Garg premise is ``macroscopic noninvasive measurability'':
a measurement can in principle distinguish which of the macroscopically
distinguishable states the system is in, with a negligible effect
on the subsequent macroscopic dynamics of the system. There have been
violations of Leggett-Garg inequalities reported, including experimentally
for superconducting qubits and single atoms \foreignlanguage{australian}{\cite{jordan_kickedqndlg2,lgexpphotonweak,Mitchell-1,lauralg,bognoon,massiveosci-1-1,NSTmunro,robens,emary-review,manushan-cat-lg}}.
A complication with the Leggett-Garg tests is the justification
of the second ``noninvasive measurability'' premise for any practical
measurement \foreignlanguage{australian}{\cite{jordan_kickedqndlg2,lgexpphotonweak,lauralg,manushan-cat-lg,legggarg}.}

In this paper, we show how a form of macroscopic realism may be tested
using Bell inequalities \cite{Bell-2}. This represents an advance
because here the second Leggett-Garg premise is replaced by the premise
of \emph{macroscopic locality} (ML), which leads to a stronger test
of macroscopic realism: Where a measurement at one location gives
one of two macroscopically distinguishable outcomes, and macroscopic
realism is assumed, then macroscopic locality implies that a measurement
made at another location cannot change the predetermined (hidden-variable)
value for the measurement at the first location. This is provided
the two measurement events are spacelike separated. In Schrodinger's
paradox, macroscopic locality implies a measurement on a second separated
system could not (instantly) change the cat from dead to alive, or
vice versa. The combined premises of MR and ML constitute the premise
of \emph{macroscopic local realism} (MLR) \cite{MLR,mlr-uncertainty,mdrmlr2}.

Specifically, we explain how the predictions of quantum mechanics
are incompatible with those of macroscopic local realism for systems
prepared in certain macroscopic entangled superposition states. To
obtain a quantifiable test for cases where macroscopically distinct
outcomes are not realistic, we define $M$-scopic local realism to
apply where the outcomes are separated by an amount of order $M$.
The important feature of the Bell tests presented in this paper is
that the outcomes of \emph{all} relevant measurements involved in
the Bell inequality  correspond to macroscopically distinct states
of the system being measured i.e. measurements only need to distinguish
between the two extreme states of a macroscopic superposition state
(a ``cat state''). The measurements do not resolve at the level
of $\hbar$. We consider two cases. In the first, measurements detect
either all of $N$ bosons in one mode, or all $N$ bosons in a second
mode. In the second case, the measurements distinguish between the
coherent states (of amplitude $\alpha$) well separated in phase space
(by, of order, $\alpha$). We determine violation of $N$-scopic
local realism for $N$ up to $20$, and of $\alpha$-scopic local
realism for $\alpha\rightarrow\infty$. The violations are possible,
because we allow nonlinear dynamics at each of the separated sites,
and consider different local time (or nonlinearity) settings.

The Bell tests of this paper differ from previous Bell tests for macroscopic
systems \cite{meso-bell-higher-spin-cat-states-bell}, including those
for superpositions of macroscopically distinct states \cite{cat-bell-wang,cat-bell},
which almost invariably require at least one measurement that resolves
microscopic outcomes, or else involve a continuous range of outcomes
\cite{mlr-uncertainty,MLR,cv-bell,mdrmlr2}. These former tests are
not in the spirit of Leggett and Garg, who considered only measurements
distinguishing the two macroscopically distinct states that form an
extreme macroscopic superposition state (so that the separation of
outcomes well exceeds the level associated with the standard quantum
limit ($\hbar$)). The results of this paper show that Bell violations
can be predicted in this macroscopic regime. To the best of our knowledge,
such tests have not been performed for $N>1$.

\emph{Bell inequality for macroscopic local realism:} For spatially
separated sites $A$ and $B$, we consider the two-qubit Bell state
\begin{eqnarray}
|\psi_{\pm,+}\rangle_{AB} & = & (|1\rangle_{A}|\pm1\rangle_{B}+|-1\rangle_{A}|\mp1\rangle_{B})/\sqrt{2}\nonumber \\
|\psi_{\pm,-}\rangle_{AB} & = & (|1\rangle_{A}|\pm1\rangle_{B}-|-1\rangle_{A}|\mp1\rangle_{B})/\sqrt{2}\label{eq:bell-states-2}
\end{eqnarray}
where $|1\rangle_{A}=|N\rangle_{a}|0\rangle_{a_{2}}$, $|1\rangle_{B}=|N\rangle_{b}|0\rangle_{b_{2}}$,
$|-1\rangle_{A}=|0\rangle_{a}|N\rangle_{a_{2}}$, $|-1\rangle_{B}=|0\rangle_{b}|N\rangle_{b_{2}}$.
Here, we consider two modes denoted $a$ and $a_{2}$, and $b$ and
$b_{2}$, at each location $A$ and $B$ respectively. $|n\rangle_{a/b}$
is the number state for the mode denoted $a/b$. We next define the
action of a hypothetical nonlinear beam splitter (NBS) at site $A$.
For an initial state $|N\rangle_{a}|0\rangle_{a_{2}}$, the state
after the hypothetical NBS interaction is 
\begin{eqnarray}
|\psi(t_{a})\rangle & = & \hat{U}_{A}|N\rangle_{a}|0\rangle_{a_{2}}\label{eq:nbsa}\\
 & = & e^{i\varphi(t_{a})}(\cos t_{a}|N\rangle_{a}|0\rangle_{a_{2}}-i\sin t_{a}|0\rangle_{a}|N\rangle_{a_{2}})\nonumber 
\end{eqnarray}
where we have introduced a unitary operator $\hat{U}_{A}$ and $t_{a}$
is the time of interaction in scaled units. A similar NBS interaction
is assumed to take place at site $B$. 
\begin{eqnarray}
|\psi(t_{b})\rangle & = & \hat{U}_{B}|N\rangle_{b}|0\rangle_{b_{2}}\label{eq:nbsb}\\
 & = & e^{i\varphi(t_{b})}(\cos t_{b}|N\rangle_{b}|0\rangle_{b2}-i\sin t_{b}|0\rangle_{b}|N\rangle_{b2})\nonumber 
\end{eqnarray}
Assuming the incoming state to be $|\psi_{+,\pm}\rangle_{AB}$, the
final state is
\begin{equation}
\hat{U}_{A}\hat{U}_{B}|\psi_{AB}\rangle=e^{i\varphi}(\cos t_{\pm}|\psi_{+,\pm}\rangle-i\sin t_{\pm}|\psi_{-,\pm}\rangle)\label{eq:bell-soln}
\end{equation}
where $t_{\pm}=t_{a}\pm t_{b}$ and $\varphi$ is a phase factor.
Defining the ``spin'' at $A$ ($B$) as $\pm1$ if the system is
detected as $|\pm1\rangle_{A}$ ($|\pm1\rangle_{B})$, the expectation
value for the spin product is $E(t_{a},t_{b})=\cos2(t_{\pm})$. Where
$N$ is large, the assumption of macroscopic local realism (MLR) will
imply the Clauser-Horne-Shimony-Holt (CHSH) Bell inequality $B\leq2$
\cite{CShim-review}, where
\begin{equation}
B=E(t_{a},t_{b})-E(t_{a},t'_{b})+E(t'_{a},t_{b}')+E(t'_{a},t_{b})\label{eq:chsh-bell}
\end{equation}
Here we note there are two choices of interaction times at each location:
$t_{a}$, $t'_{a}$ at $A$, and $t_{b}$, $t_{b}'$ at $B$. This
inequality is derived assuming that before the measurement at the
selected time $t$, the system is in one or other states described
by $|\pm1\rangle$ at each site (MR), and that there is no nonlocal
effect changing the state due to the measurement at the other location
(ML). For $N$ large, the qubits $|\pm1\rangle_{A/B}$ correspond
to macroscopically distinct outcomes for all choices of $t_{a}$ and
$t_{b}$, and the violation of the Bell inequality will falsify MLR.
The solution for $E(t_{a},t_{b})$ will violate the inequality for
suitable choices of $t_{a}$ and $t_{b}$ \cite{CShim-review}. 
\begin{figure}[t]

\begin{centering}
\includegraphics[width=0.52\columnwidth]{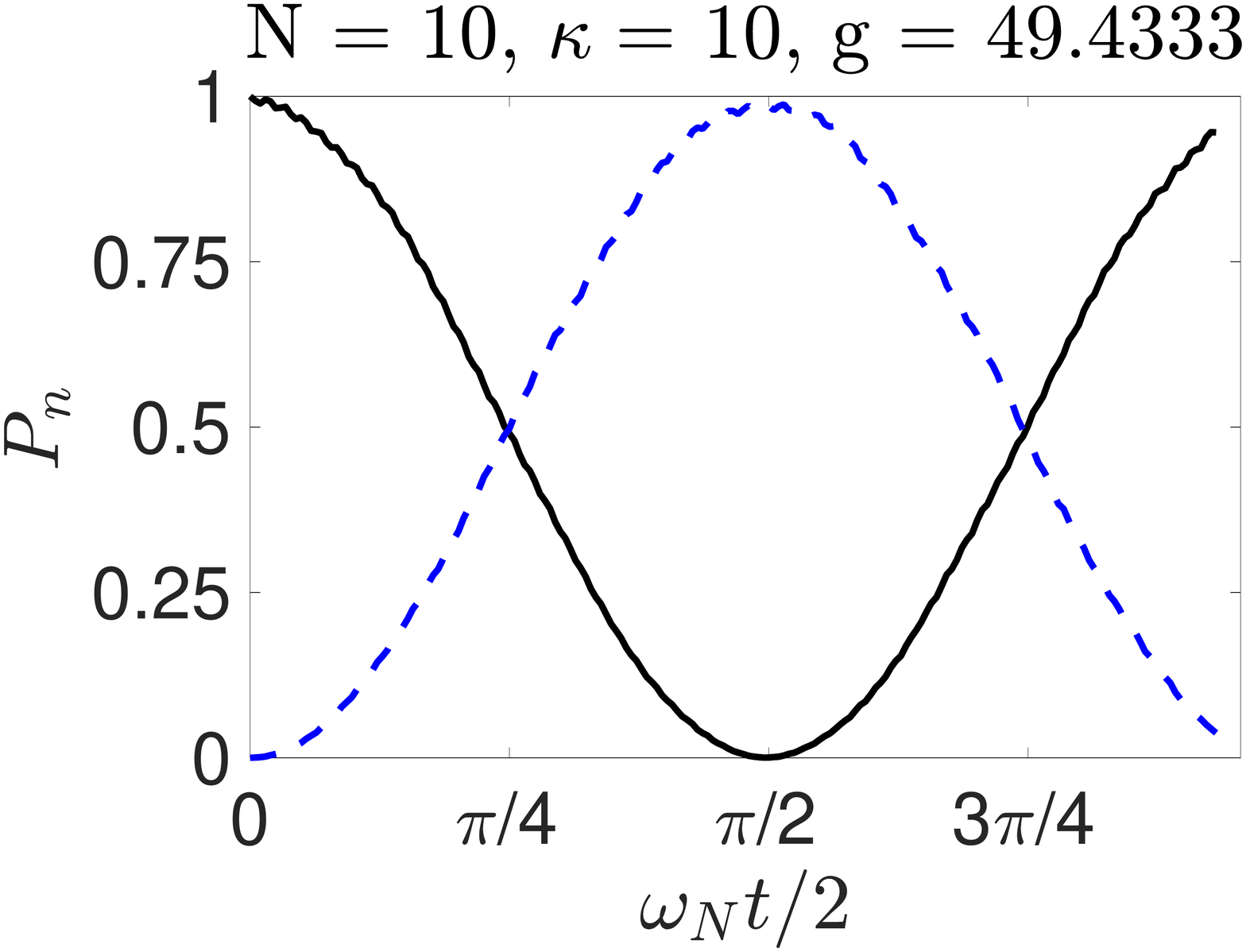}\includegraphics[width=0.52\columnwidth]{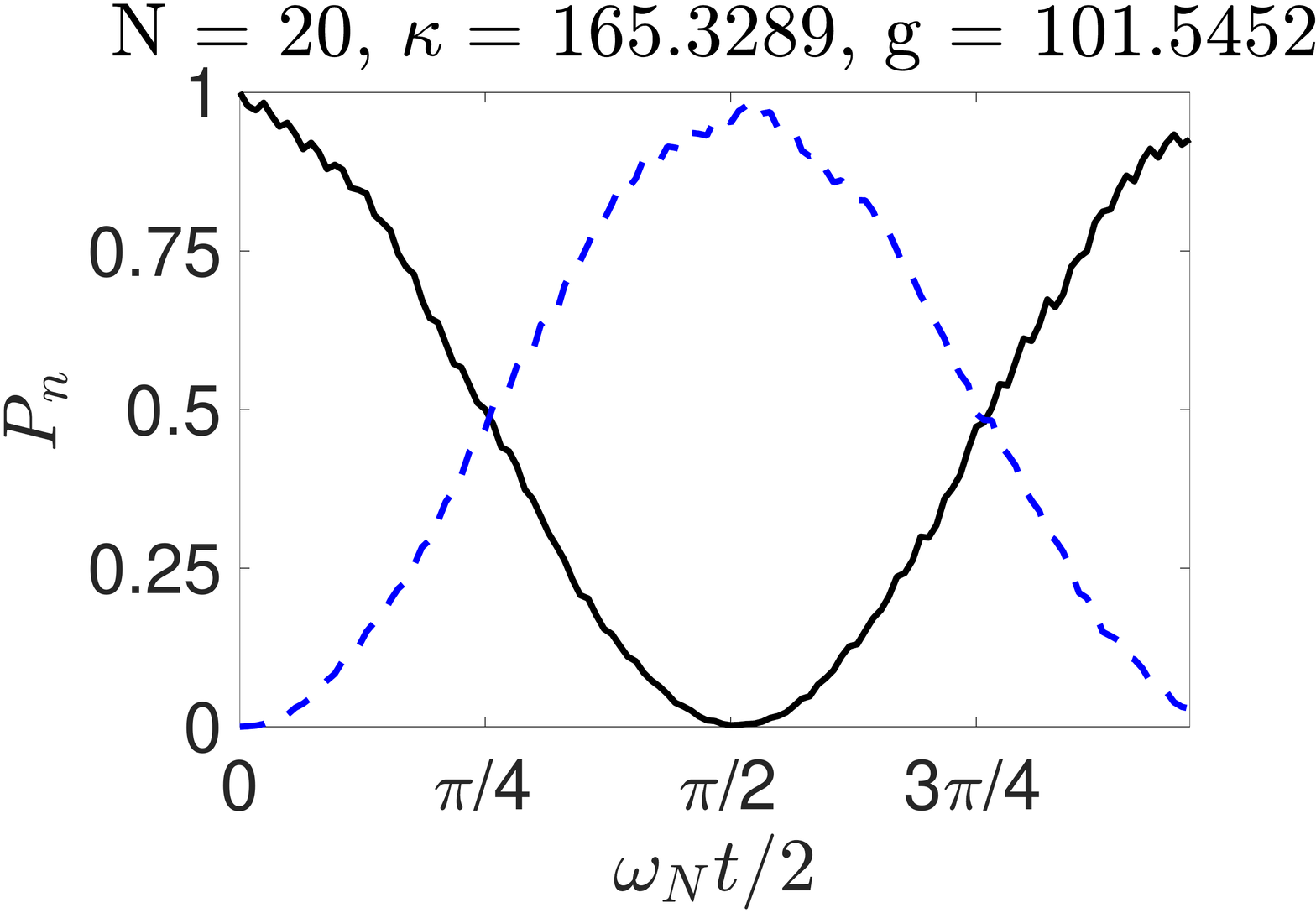}
\par\end{centering}
\caption{Solutions for the Hamiltonian $H_{NL}^{(A)}$ after a time $t$ with
initial state $|N\rangle_{a}|0\rangle_{a_{2}}$.\textcolor{red}{{} }$P_{N}$
(black solid line) is the probability for all $N$ bosons to be in
mode $a$;  $P_{0}$ (blue dashed line) is the probability for all
$N$ bosons to be in mode $a_{2}$. The parameters identify regimes
optimal, or near-optimal, for the nonlinear beam splitter interaction
eq. (\ref{eq:nbsa}), where $P_{N}+P_{0}\sim1$ and $P_{N}\sim\cos^{2}\omega_{N}t$.
\textcolor{red}{}Solutions for $N=1$ (all $\kappa,$$g$); \textcolor{black}{$N=2$,
$\kappa=1$, $g=30$; $N=5$, $\kappa=20$, $g=333.333$; and }$N=7$,
$\kappa=18.23$, $g=47.85$ are similar to the left plot.\textcolor{red}{}}
\end{figure}

\emph{Nonlinear beam splitter (NBS) for $N$ bosons: }The above is
a straighforward extension of Bell's work, except for the nontrivial
complication that it needs to be shown that the hypothetical NBS interaction
can be predicted in quantum mechanics, to a sufficient degree that
allows the violation of the Bell inequality. To do this, we consider
at $A$ two incoming fields ($a$ and $a_{2}$), which interact according
to the nonlinear Josephson Hamiltonian \cite{josHam-lmg,josHam-collett-steel}
\begin{equation}
H_{NL}^{(A)}=\kappa(\hat{a}^{\dagger}\hat{a}_{2}+\hat{a}\hat{a}_{2}^{\dagger})+g\hat{a}^{\dagger2}\hat{a}^{2}+g\hat{a}_{2}^{\dagger2}\hat{a}_{2}^{2}\label{eq:ham}
\end{equation}
Here, $\hat{a}$, $\hat{a}_{2}$ are the boson operators for the corresponding
fields $a$, $a_{2}$, modelled as single modes. A similar interaction
$H_{NL}^{(B)}$ is assumed for the fields $b$ and $b_{2}$ at $B$.
Such an interaction can be achieved with Bose-Einstein condensates
(BEC) or superconducting circuits \cite{oberoscexp,pol.,superconducting-nonlinear,superfluid,two-well-cm,carr-two-well,josHam-collett-steel,josHam-lmg,bognoon,lauralg}.
For certain choices of $g$ and $\kappa$, we find that the interaction
(\ref{eq:ham}) acts as a nonlinear beam splitter, where the input
$|N\rangle_{a}|0\rangle_{a_{2}}$ after a time $t$ gives, to a good
approximation, the output of eqn (\ref{eq:nbsa}) (Figure 1). We
introduce a scaled time \textcolor{black}{$t_{a}=\omega_{N}t$ where
$\omega_{N}=2g\frac{N}{\hbar(N-1)!}\left(\frac{\kappa}{g}\right)^{N}$
\cite{carr-two-well}.}\textcolor{red}{{} }

\emph{$N$-scopic Bell tests for $N$ bosons:} It now remains to determine
whether the realisation of the NBS (which is never exact) can \emph{actually}
allow a violation of the Bell inequality. We first examine a specific
method of preparation of (\ref{eq:bell-states-2}) \cite{noon-cond-pryde-white,reid-walls,Ou-mandel,laura-paper,int-nonlocality-bS}.
We consider that the two separated modes $a$ and $b$ are prepared
at time $t=0$ in the NOON state $|\psi\rangle_{ab}=\frac{1}{\sqrt{2}}(|N\rangle_{a}|0\rangle_{b}+e^{i\vartheta}|0\rangle_{a}|N\rangle_{b})$
\cite{opticalNOON-1,heralded-noon,laura-paper,herald-noon} and that
the modes $a_{2}$ and $b_{2}$ are prepared similarly, in the NOON
state $\frac{1}{\sqrt{2}}(|0\rangle_{a_{2}}|N\rangle_{b_{2}}+e^{-i\pi/2}|N\rangle_{a_{2}}|0\rangle_{b_{2}})$
(Figure 2). Assuming optimal NBS parameters, the final state is
$|\psi_{f-}\rangle=\hat{U}_{A}\hat{U}_{B}|\psi\rangle_{ab}|\psi\rangle_{a_{2}b_{2}}$.
For an alternative method using the NBS interaction, refer to the
Supplemental Materials \cite{supmat}. We find ($\phi_{-}$ is\textcolor{black}{{}
a phase factor)}
\begin{eqnarray}
|\psi_{f-}\rangle & = & \frac{e^{i\phi_{-}}}{\sqrt{2}}\Bigl(\cos t_{-}(|\psi_{-,-}\rangle{\color{black}{\color{black}-}}i\sin t_{-}|\psi_{+,-}\rangle\Bigl)\nonumber \\
 &  & +\frac{1}{\sqrt{2}}|\varphi_{-}\rangle_{2N}\label{eq:soln1}
\end{eqnarray}
$|\varphi_{-}\rangle_{2N}$ are states with all $2N$ bosons at
site $A$, or all $2N$ bosons at site $B$.\textcolor{red}{}\textcolor{blue}{}

\begin{figure}[t]
\vspace{-2cm}
\includegraphics[width=1\columnwidth]{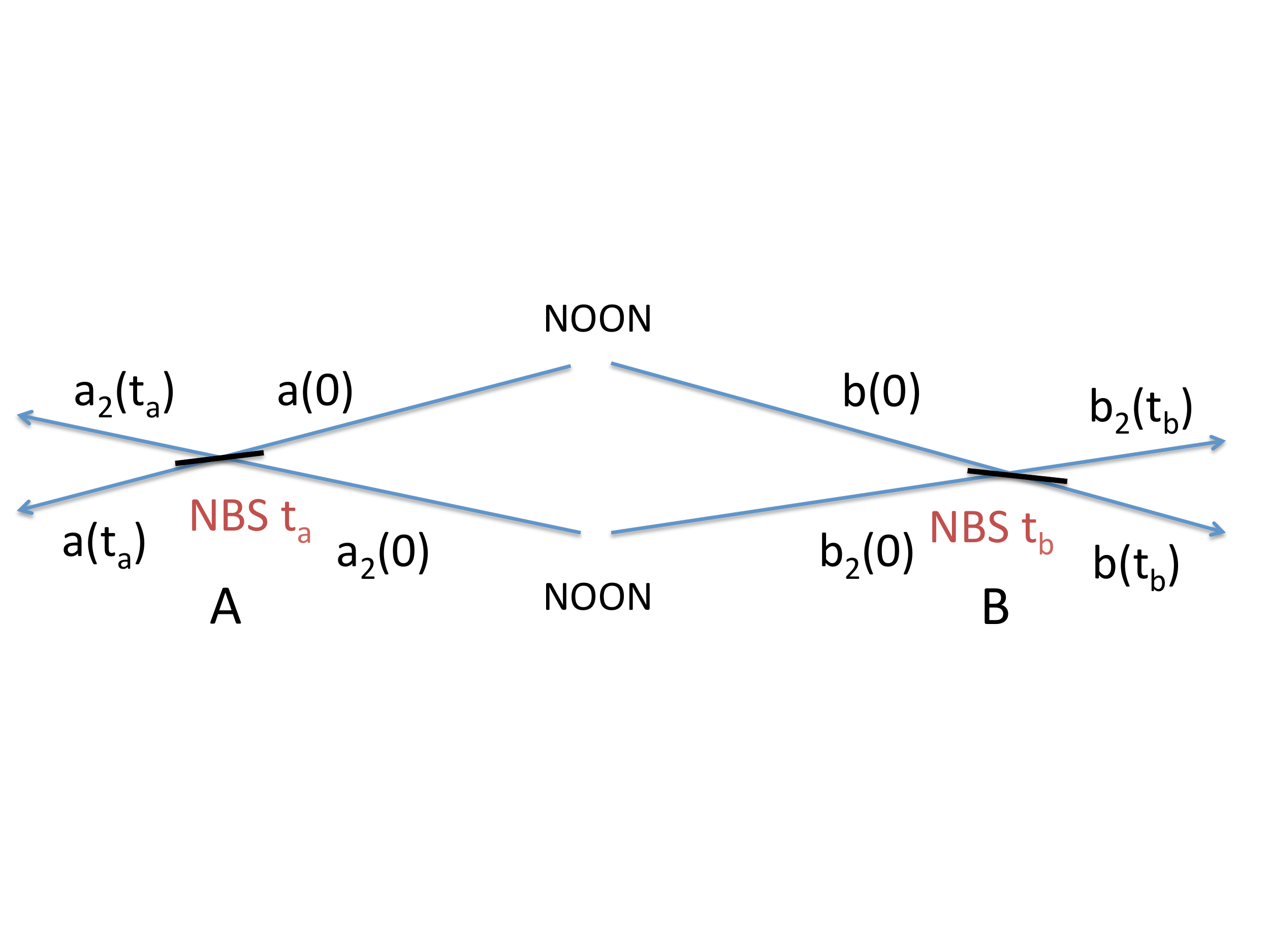}\vspace{-2cm}

\caption{The macroscopic entangled state (\ref{eq:bell-states-2}) can be generated
conditionally by interfering two NOON states. The $t_{a}$ and $t_{b}$
are time settings selected for the nonlinear beam splitter (NBS) at
each site, $A$ and $B$.}
\end{figure}

To test $N$-scopic local realism, the mode numbers at the final outputs
$a(t_{a})$, $a_{2}(t_{a})$, $b(t_{b})$ and $b_{2}(t_{b})$ are
measured (Figure 2), for a given setting of the times $t_{a}$ and
$t_{b}$ at each site. The measurement events at $A$ and $B$ are
spacelike separated if the distance between them is sufficiently great,
taking into account the times $t_{a}$ and $t_{b}$ required for the
NBS interactions. \textcolor{black}{At $A$, we denote by $+$ the
outcome of detecting $N$ bosons at location $a$, }\textcolor{black}{\emph{and}}\textcolor{black}{{}
$0$ bosons at $a_{2}$. A similar outcome $+$ is defined for $B$.
(The state }$|\varphi_{-}\rangle_{2N}$ \textcolor{black}{thus becomes
irrelevant). We define the joint probability $P_{++}$ for the outcome
$+1$ at both sites $A$ and $B$. We also specify $P_{+}^{A}$ as
the marginal probability for the outcome $+$ at $A$, and $P_{+}^{B}$
as the marginal probability for the outcome $+$ at $B$. }At each
site $A$ and $B$, observers independently select a time of evolution
$t_{a}$ and $t_{b}$ for the NBS interaction.

It is evident from the expression (\ref{eq:soln1}) that the outcomes
of a measurement of mode number at each detector are \textcolor{black}{always
one of $0$, $N$ or $2N$} (Figure 3), which are macroscopically
distinguishable as $N\rightarrow\infty$.  The assumption of $N$-scopic
LR \textcolor{black}{(which becomes MLR as $N\rightarrow\infty$)}
thus implies the validity of a local hidden variable theory, where
the system at each site is predetermined to be in one of the states
with mode number $0$, $N$, or $2N$. The Clauser-Horne (CH) Bell
inequality $S\leq1$ \textcolor{black}{\cite{CShim-review} is predicted
to hold for such a local hidden variable theory, where \cite{laura-paper}}
\begin{eqnarray}
S & = & \frac{P_{++}(t_{a},t_{b})-P_{++}(t_{a},t'_{b})+P_{++}(t_{a}',t{}_{b})+P_{++}(t'_{a},t'_{b})}{P_{+}^{A}(t_{a}')+P_{+}^{B}(t_{b})}\nonumber \\
\label{eq:ch-bell-ineq}
\end{eqnarray}
\textcolor{black}{Assuming }\textcolor{black}{\emph{ideal}}\textcolor{black}{{}
nonlinear beam splitters, the state $|\psi_{f-}\rangle$ gives $P_{++}=\frac{1}{4}\sin^{2}(t_{a}-t_{b})$
and $P_{+}^{A}=P_{+}^{B}=\frac{1}{4}$. For $t_{a}=0$, $t_{a}'=2\varphi$,
$t_{b}=\varphi$ and $t_{b}'=3\varphi$,  the quantum state $|\psi_{f-}\rangle$
of (\ref{eq:soln1}) predicts $S=\frac{3\sin^{2}(\varphi)-\sin^{2}(3\varphi)}{2}$}.
$S$ maximizes at $S=1.207$ for $\varphi=\pi/16$, giving a violation
of the CH Bell inequality.\textcolor{black}{}
\begin{figure}[t]

\includegraphics[width=0.8\columnwidth]{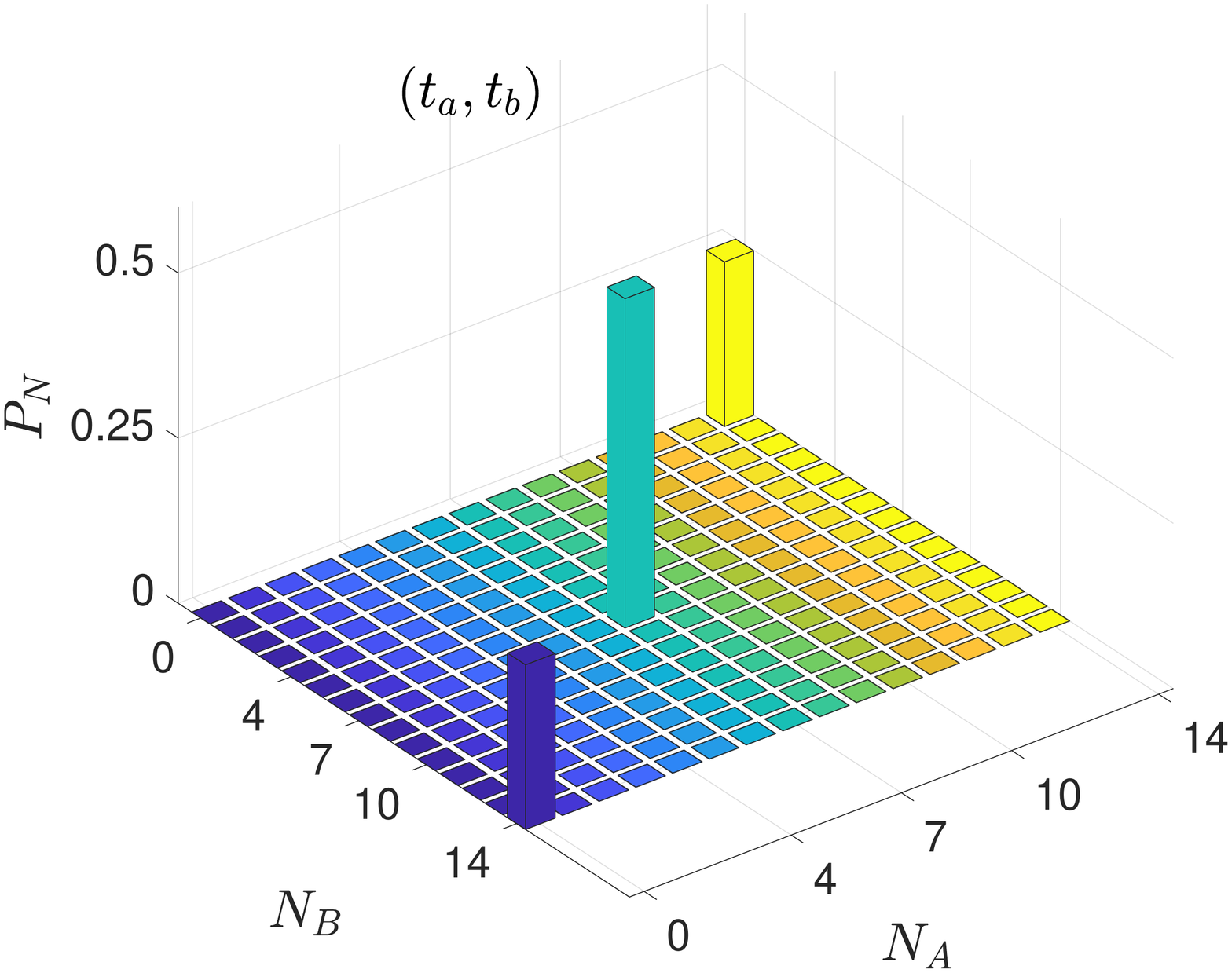}

\includegraphics[width=0.5\columnwidth]{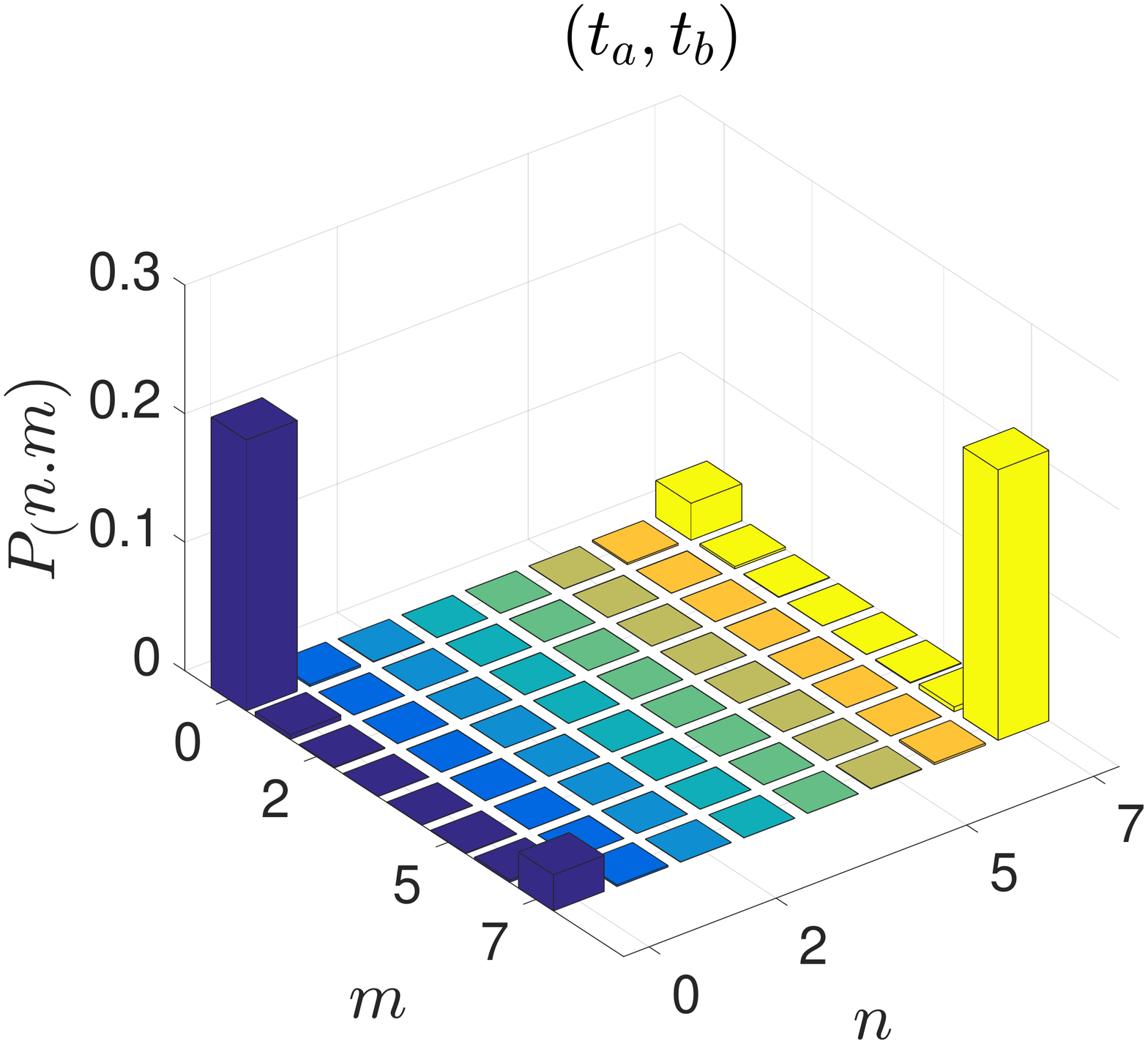}\includegraphics[width=0.5\columnwidth]{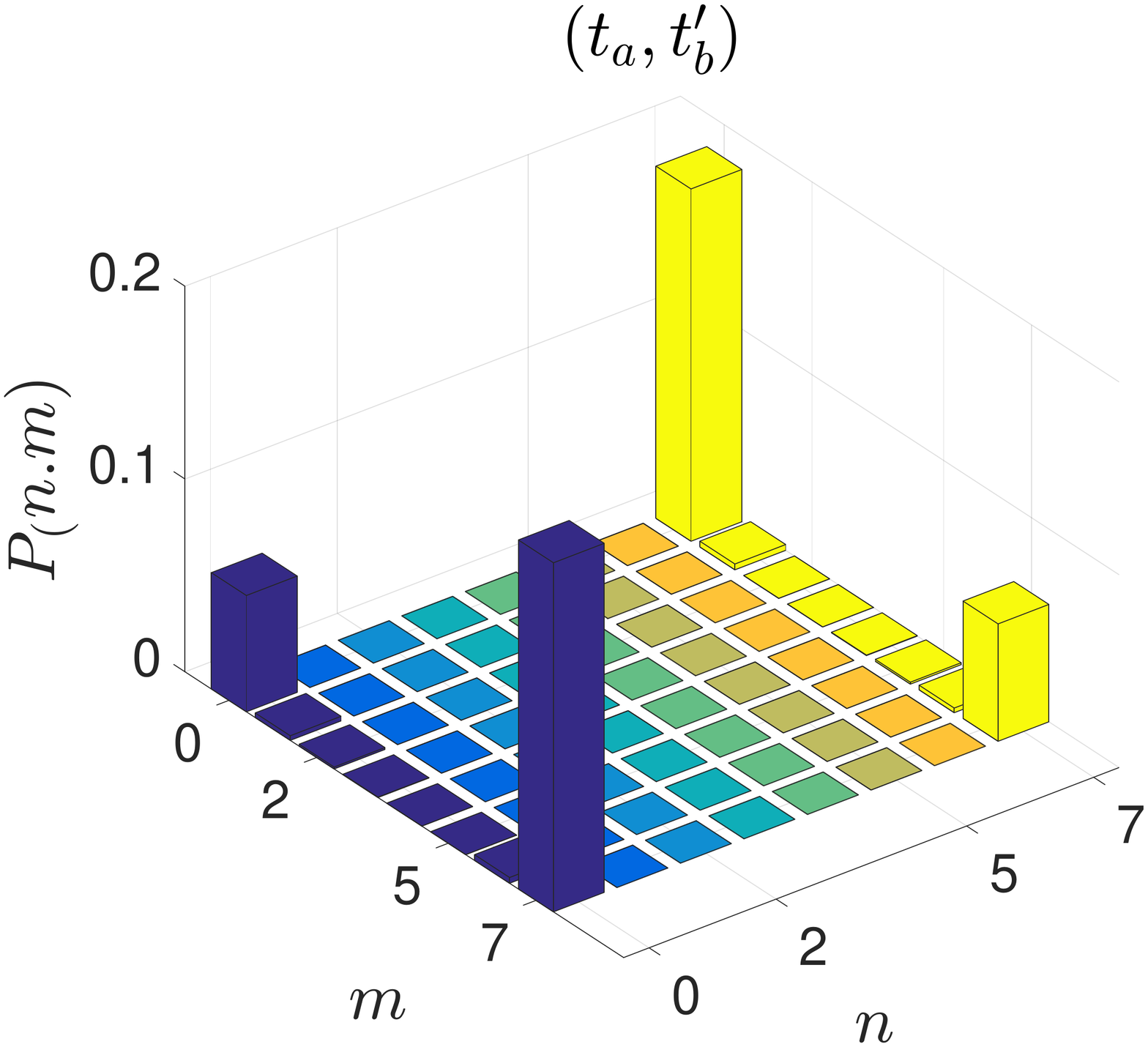}

\caption{Top: Probability distribution $P_{N}$ for the joint detection of
$N_{A}$ and $N_{B}$ bosons at the sites $A$ and $B$ (refer Fig.
2). Here $t_{a}$, $t_{b}$, \textcolor{black}{$t_{a}'$, $t_{b}'$
and $\varphi$ are specified in the text. }$N=7$, $\kappa=18.23$,
$g=47.85$\textcolor{black}{. The distribution is unchanged for settings
$(t'_{a},t_{b})$, $(t'_{a},t'_{b})$}\textbf{\textcolor{black}{{} }}\textcolor{black}{and
}\textbf{\textcolor{black}{$(t_{a},t'_{b})$}}\textcolor{black}{.}\textbf{\textcolor{black}{{}
}}\textcolor{black}{Lower:}\textcolor{red}{{} }\textcolor{black}{T}he
joint probability $P(n,m)$ of detecting $n$ bosons in mode $a$
and $m$ bosons in mode $b$, \emph{and} $N$ bosons in total at site
$A$. \textcolor{blue}{}\textcolor{red}{}\textcolor{black}{The
figures for settings  $(t'_{a},t_{b})$, $(t'_{a},t'_{b})$ are identical
to those of $(t_{a},t_{b})$.}\textbf{\textcolor{red}{{} }}\textcolor{red}{}\textcolor{black}{Similar
plots are obtained for all parameters given in Fig. 1.}}
\end{figure}

\begin{figure}[t]

\textcolor{red}{\includegraphics[width=0.5\columnwidth]{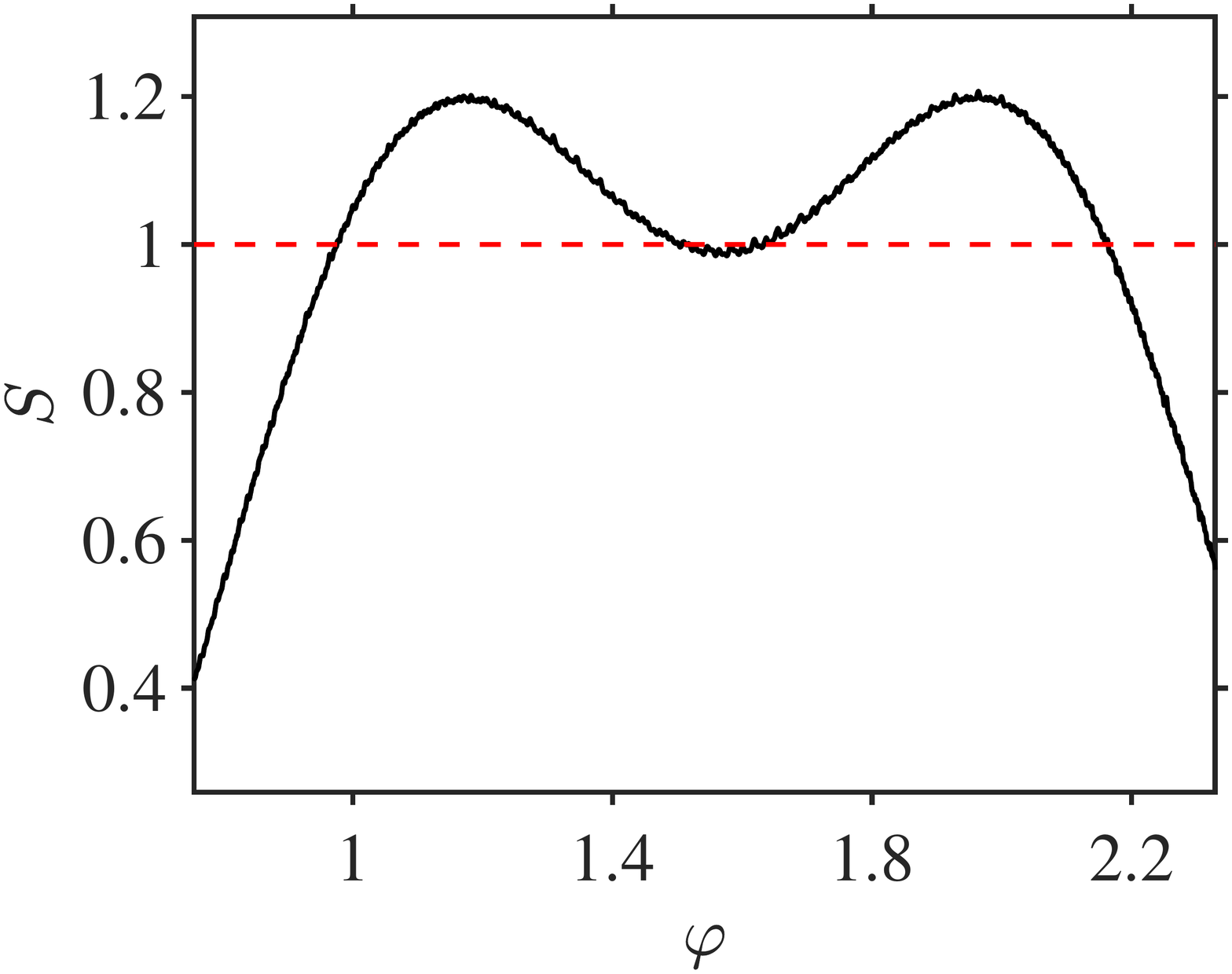}}\includegraphics[width=0.54\columnwidth]{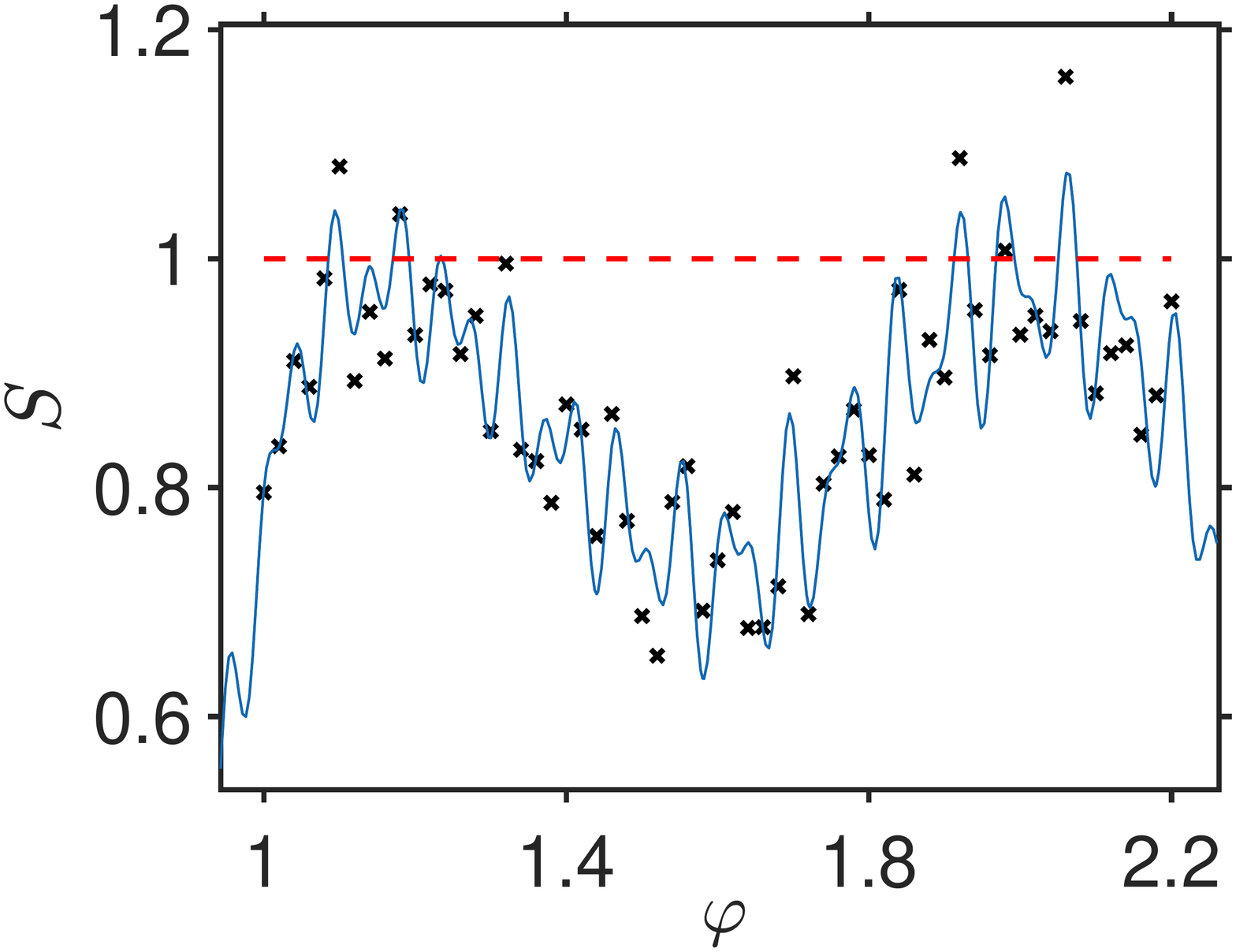}

\caption{Violation of the CH-Bell inequality (\ref{eq:ch-bell-ineq}) is obtained
when $S>1$. Left graph is for $N=10$, $g=49.433$, $\kappa=10$.
Plots for $N=1$ (all $\kappa$ and $g$); $N=2$, $\kappa=1$, $g=30$;
\textcolor{black}{$N=5$, $\kappa=20$, $g=333.333$; and }$N=7$,
$\kappa=18.23$, $g=47.85$ which give ideal two-state oscillatory
behaviour (Fig. 1) are almost indistinguishable. \textcolor{red}{}Where
the values are not quite optimal,  rapid oscillations of small amplitude
appear. This is shown in the right graph for \textcolor{red}{}
$N=20$, $\kappa=165$ and $g=101$.\textcolor{red}{}}
\end{figure}

\emph{}In practice, the ideal regime giving the precise solution
(\ref{eq:nbsa}) for the nonlinear beam splitters is unattainable,
for $N>1$, since probabilities for other than  $0$ or $N$ bosons
in each mode are not precisely zero. In Figures 3 and 4, we present
\emph{actual }predictions for $S$, using the Hamiltonian $H_{NL}$.
For large $gN/\kappa$, where care is taken to optimize for the NBS
regime given by (\ref{eq:nbsa}), the Bell violations are predicted,
as shown in Figure 4. To test N-scopic LR, one requires to establish
that the outcomes of mode number are distinct by $N$, for \emph{each}
of the joint probabilities $P(t_{a},t_{b})$ comprising $S$. Figure
3 highlights this feature in the optimal parameter space. The probabilities
for results other than $0$ and $N$ (and $2N$, for Figure 3a) are
negligible (and can rigorously be shown to have no effect on the violation,
using the methods of \cite{eric_marg,legggarg}).\textcolor{red}{}

\emph{Macroscopic Bell tests using cat-states:} To examine macroscopic
behaviour, we consider the Bell cat-state
\begin{equation}
|\psi\rangle=N(|+\rangle_{a}|+\rangle_{b}-|-\rangle_{a}|-\rangle_{b})\label{eq:cat}
\end{equation}
where $|+\rangle_{a}=-e^{i\pi/6}(|e^{i\pi/3}\alpha\rangle_{a}+|e^{-i\pi/3}\alpha\rangle_{a})/\sqrt{2}$,
$|-\rangle_{b}=ie^{-i\pi/6}(|-e^{i\pi/3}\beta\rangle_{b}+|-e^{-i\pi/3}\beta\rangle_{b})/\sqrt{2}$,
$|-\rangle_{a}=|-\alpha\rangle$, $|+\rangle_{b}=-i|\beta\rangle$
and $|\alpha\rangle$, $|\beta\rangle$ are coherent states for two
modes labelled $a$ and $b$. $N$ is a normalisation constant. Since
we consider $\alpha,\beta$ to be real and $\alpha$, $\beta\rightarrow\infty$,
$|\pm\rangle_{a}$ become orthogonal, and similarly $|\pm\rangle_{b}$.
This then corresponds to the state $|\psi_{+,-}\rangle_{AB}$ of
(\ref{eq:bell-states-2}), where $|\pm\rangle_{A/B}=|\psi_{\pm}\rangle_{a/b}$.

To realise a nonlinear interaction (NBS), we propose at each site
the nonlinear Kerr interaction \textcolor{black}{{} $H_{NL}^{(A/B)}=\Omega\hat{n}_{A/B}^{2}$
where $\hat{n}_{A}=\hat{a}^{\dagger}\hat{a}$ and $\hat{n}_{B}=\hat{b}^{\dagger}\hat{b}$
\cite{yurke-stoler,wrigth-walls-gar,kirchmair,collapse-revival-bec}.}
For systems prepared in coherent states $|\alpha\rangle$ the interaction
after certain times leads to the formation of cat-states where (for
large $\alpha$) the system is in a superposition of macroscopically
distinct coherent states in phase space.\textcolor{black}{{} At any
time, one can perform quadrature phase amplitude measurements $\hat{X}_{A}={\color{red}{\color{blue}{\color{black}\frac{1}{\sqrt{2}}}}}(\hat{a}+\hat{a}^{\dagger})$
and $\hat{X}_{B}={\color{red}{\color{blue}{\color{black}\frac{1}{\sqrt{2}}}}}(\hat{b}+\hat{b}^{\dagger})$
at each site. The ``spin'' result $\hat{s}$ is taken to be $+1$
if the result for such a measurement is $>0$, and $-1$ otherwise.
A state with outcome $\pm1$ is denoted $|\pm\rangle$. }At time\textcolor{black}{{}
$t'_{A}=\pi/(3\Omega)$, if the initial state is $|\alpha\rangle$,
}the state at $A$ is \textcolor{black}{\cite{manushan-cat-lg}}
\begin{eqnarray}
|\psi(t'_{A})\rangle & = & -i{\color{blue}{\normalcolor \sqrt{\frac{1}{3}}|-\rangle_{a}+\sqrt{\frac{2}{3}}|+\rangle_{a}}}\label{eq:23}
\end{eqnarray}
For an initial state $|\beta\rangle$, the state at time $t'_{b}=2\pi/(3\Omega)$
is
\begin{eqnarray}
{\normalcolor |\psi(t'_{B})\rangle} & {\normalcolor =} & \sqrt{\frac{1}{3}}|+\rangle_{b}-i\sqrt{\frac{2}{3}}|-\rangle_{b}\label{eq:78}
\end{eqnarray}
\textcolor{black}{We select $t_{A}=0$ and $t'_{A}=\pi/(3\Omega)$,
and $t_{B}=0$ and }$t'_{B}=2\pi/(3\Omega)$. The final state after
the Kerr dynamics gives (as $\alpha=\beta\rightarrow\infty$): $E(t_{A},t{}_{B})=1$;
$E(t_{A},t'{}_{B})=-1/3$; $E(t'_{A},t{}_{B})=1/3$; $E(t'_{A},t'{}_{B})=7/9$.
This implies violation of the Bell inequality (\ref{eq:chsh-bell})
with $B=2.44$ \cite{supmat}.

Figure 5 gives the complete predictions for arbitrary $\alpha$, $\beta$,
accounting for the full effect of nonorthogonality of the coherent
states. The outcomes of measurements of $\hat{s}$ (the sign of $\hat{X}$)
are macroscopically distinct\emph{, }corresponding to macroscopically
distinguishable states\emph{ }in phase space, for \textcolor{black}{
}\textcolor{black}{\emph{all}}\textcolor{black}{{} of the choices of
time-settings, as $\alpha,$$\beta$ becomes large }(Figure 5a and
\cite{supmat}). Violations of $\alpha$-scopic local realism are
predicted for all $\alpha=\beta>2$ (Figure 5b) \cite{supmat}.
\begin{figure}[t]

\includegraphics[width=0.5\columnwidth]{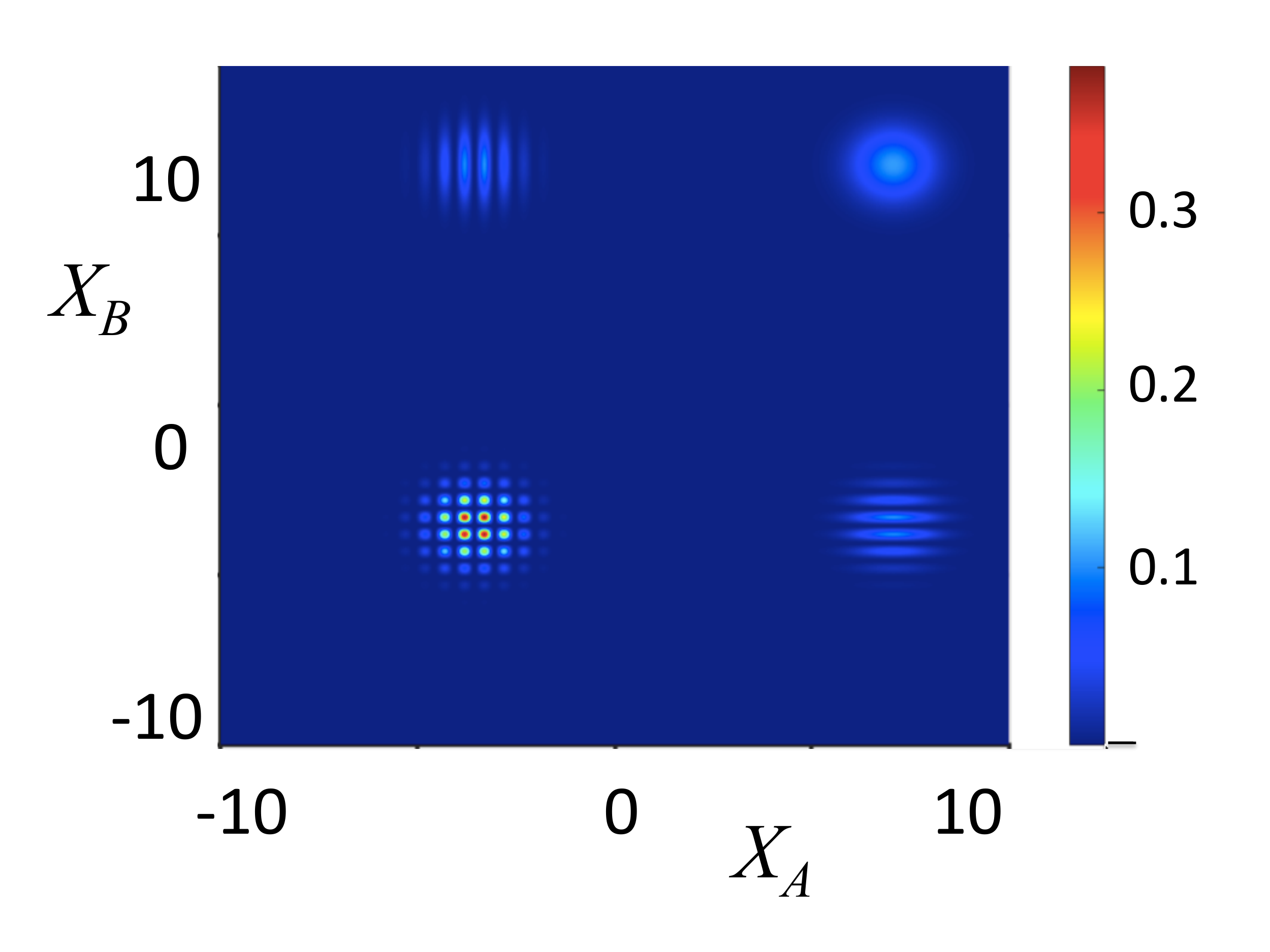}\includegraphics[width=0.5\columnwidth]{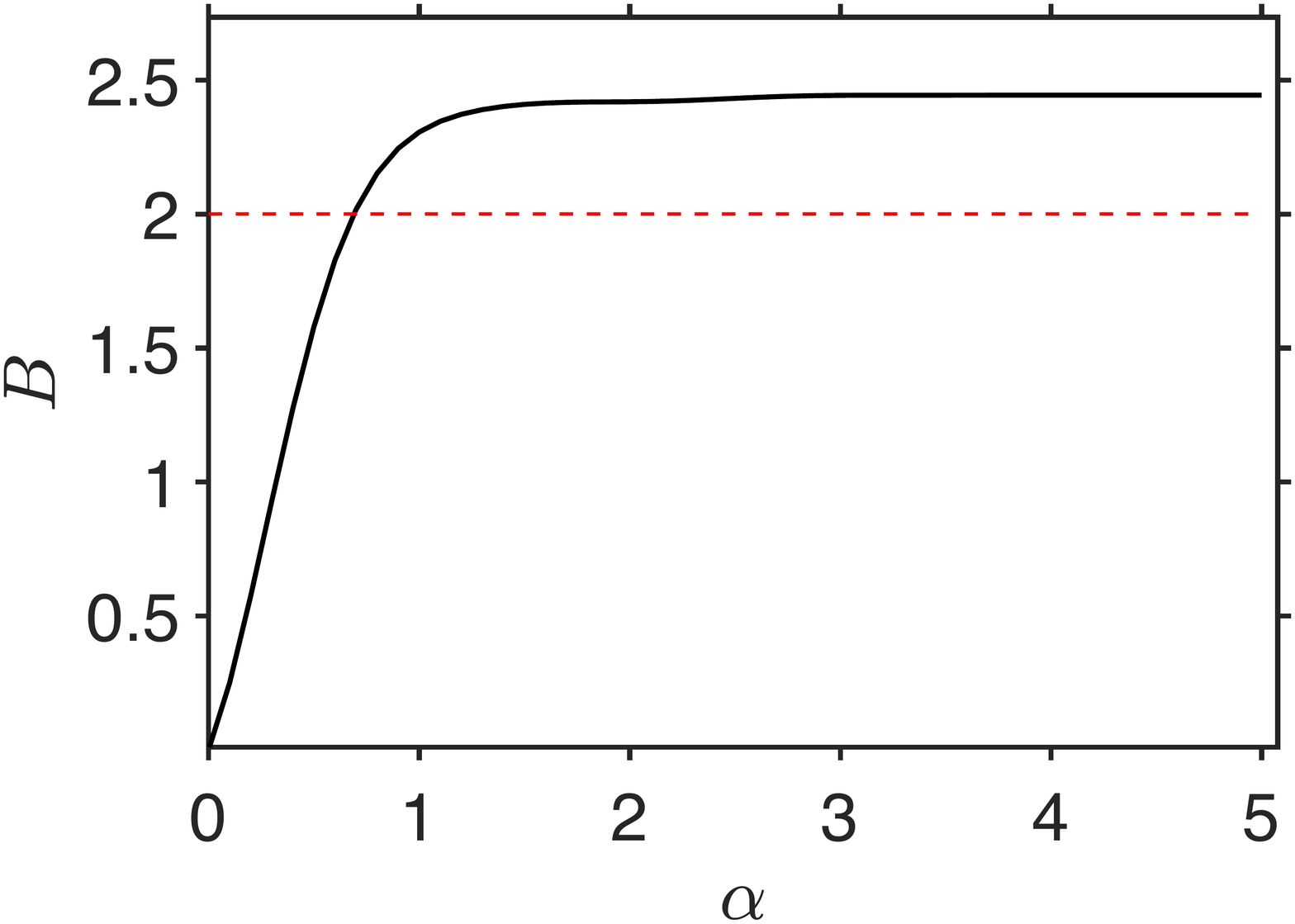}

\caption{Left: A contour plot for the joint probability distribution of the
quadrature phase amplitudes $X_{A}$ and $X_{B}$ at times $t_{a}=\pi/3\Omega$
and $t_{b}=0$. The four outcomes depicted are macroscopically distinct
for $\alpha$ large. Here $\alpha=\beta=5$. The distributions for
the remaining 3 pairs of times are similar \cite{supmat}. Right:
The corresponding violation $B>2$ of the Bell inequality eq. (\ref{eq:chsh-bell})
for arbitrarily large $\alpha=\beta$.\textcolor{red}{ }}
\end{figure}

\emph{Conclusion: }We have argued that violation of the Bell inequality
falsifies MLR, because the outcomes measured at location $A$ ($B$)
are macroscopically distinct:  A critic might claim differently.
They may argue that a microscopic nonlocal quantum effect is translated
by the dynamics (which occurs over a time $t_{a/b}$) into a macroscopic
effect, which is then registered by the detectors. This criticism
could be further explored \cite{supmat}.

Preparing entangled macroscopic superposition states where $A$ and
$B$ are spatially separated is a challenge. This is addressed if
the initial NOON state of Figure 2 has separated modes (for $N=2$,
the Hong-Ou-Mandel effect might be useful \cite{heralded-noon}).For
a Rb BEC, the timescales required for the nonlinear beam splitter
become inaccessibly long \cite{oberoscexp,carr-two-well}.  The nonlinear
beam splitter is however likely achievable using superconducting circuits
to obtain high nonlinearities \cite{bellcatexp,superconducting-nonlinear}.
In fact, a two-mode cat-state similar to (\ref{eq:cat}) has been
generated \cite{cat-bell-wang} (although without spatial separation)
and the dynamics of (\ref{eq:23}) and (\ref{eq:78}) realized for
BECs and microwave fields \cite{collapse-revival-bec,kirchmair}.
Noting that macroscopic realism (MR) is suggestive of the validity
macroscopic locality (ML) \cite{det-bell}, an experimental test,
even if without spatial separation, for moderate $N$ or $\alpha$,
would be of interest.
\begin{acknowledgments}
We thank the Australian Research Council Discovery Project Grants
schemes under Grant DP180102470. \textcolor{black}{M.D.R thanks the
hospitality of the Institute for Atomic and Molecular 1009 Physics
(ITAMP) at Harvard University, supported by the NSF.}
\end{acknowledgments}

\end{document}